# Integrative Mobility Model for Grain-Boundary-Limited Transport In Thermoelectric Compounds

Gbadebo Taofeek Yusuf[1,4], Sukhwinder Singh[2*], Alexandros Askounis[1], Zlatka Stoeva[3] and Fideline Tchuenbou-Magaia[1*]

[1] Energy and Green Technology Research Group, Centre for Engineering Innovation and Research, University of Wolverhampton, Telford Innovation Campus, Telford TF2 9NT, UK

[2] Magnetics and Materials Research Group, School of Engineering, Cardiff University, Cardiff, UK, CF24 3AA

[3]DZP Technologies Limited, Cambridge CB4 2HY, United Kingdom

[4] Department of Science Laboratory Technology (Physics), Osun State Polytechnic, Iree Nigeria

**Abstract**: Grain-boundary-limited charge transport remains a key bottleneck in polycrystalline thermoelectric materials, where reduced carrier mobility degrades electrical conductivity and suppresses power factor. This work presents a semi-empirical mobility model that integrates three dominant grain-boundary mechanisms: (i) weighted mobility linked to carrier effective mass and concentration, (ii) thermionic emission across grain-boundary barriers ($\Phi_{GB}$)and (iii) geometric suppression due to finite mean free path ($\ell$). The developed model was validated using a diverse set of polycrystalline thermoelectric materials including $Bi_2Te_3$, PbTe, $Mg_2Si$, and SnSe and demonstrated excellent agreement with experimental data ($R^2$ = 0.93–0.99) and yielding physically consistent parameters: $\Phi_{GB} \approx$ 0–0.15 eV and $\ell \approx$ 15–60 nm. The model captures non-monotonic mobility trends arising from the interplay between barrier activation and phonon scattering. To demonstrate its predictive capability, the model was applied to explore the strategies in Al-doped ZnO thermoelectric materials. This study show that combined grain-boundary passivation – reducing $\Phi_{GB}$ from 0.15 eV to 0.05 eV and moderate grain growth- extending $\ell$ from 5 nm to 25 nm- could improve the power factor by approximately 6-fold (from ~4 to ~26 mW/m·K²), while raising the electronic quality factor B by nearly 7-fold (from ~0.15 to over 1.0 × $10^{-3}$ m²/V·s·kg$^{3/2}$), approaching values seen in leading chalcogenide thermoelectrics. The model offers a transparent and practical framework for grain-boundary engineering strategies in oxide-based thermoelectric materials.

## 1. Introduction

Thermoelectric (TE) materials convert heat into electricity and vice versa and gives prospects for waste-heat recovery and solid-state cooling. TE materials required a balance between the Seebeck coefficient (S), electrical conductivity ($\sigma$) and total thermal conductivity (k) to achieve higher figure of merit (ZT=$S^2\sigma/ke + kL$ [1]. Traditional TE compounds like bismuth telluride ($Bi_2Te_3$)[2], [3] and lead telluride (PbTe)[4] achieve substantial performance at specific temperature ranges but face limitations such as costly or toxic elements and inherent thermal conductivity (k)[5]. This has necessitated research into earth-abundant alternatives (e.g. ZnO, $Mg_2Si$, SnSe) and microstructural engineering to push performance limits[5], [6], [7].

Polycrystalline samples are far more practical to produce than single crystals, but grain boundaries in polycrystals is a major challenge as they disrupt charge carrier transport which reduces mobility ($\mu$) and electrical conductivity ($\sigma$) [8]. On the other hand, grain boundaries can strongly scatter phonons to reduce lattice thermal conductivity ($k_l$) in line with the "phonon-glass, electron-crystal (PGEC)" concept for TE materials[9], [10]. Thus, an inherent trade-off arises introducing a high density of grain boundaries or employing nanostructuring can effectively suppress lattice thermal conductivity ($k_{lat}$), realizing a "phonon-glass" behaviour. However, this often comes at the expense of electrical conductivity ($\sigma$), leading to a detrimental "electron-glass" scenario due to enhanced carrier scattering at interfaces and grain boundaries [9]. Optimizing this trade-off is a core problem in TE material design. Prior studies have shown that polycrystalline SnSe has a record high ZT in single-crystal form but suffers much lower $\mu$ in its polycrystalline form due to grain-boundary scattering [5], [11]. Grain boundary scattering is identified as the dominant mobility-limiting mechanism in polycrystalline SnSe that is absent in single crystals [7]. Similarly, ZnO is a wide-bandgap metal oxide thermoelectric material whose low- and mid-temperature $\sigma$ curtailed by grain boundary barriers [5]. However, single-crystal ZnO maintains higher $\mu$ until phonon scattering dominates at high T [12]. In $Bi_2Te_3$-based alloys (the premier TE at ambient temperatures), grain-boundary engineering approach is used to reduce thermal conduction, but too many or poorly controlled grain boundaries can degrade the inherently high carrier mobility down to unacceptable levels [13].

Historically, models for charge transport in polycrystalline semiconductors date back to Petritz (1956) and (1975), who treated grain boundaries as charge trapping sites that create energy barriers for carriers [14], [15]. These models predict an exponential temperature-activated behaviour for $\sigma$ or $\mu$ when grain barriers are significant a hallmark seen in lightly doped polycrystalline silicon and oxides, where $\mu$ increases exponentially with temperature at low T [14], [15]. However, such early models often assumed simplistic scenarios (e.g. completely depleted grain boundaries, or negligible intragrain scattering) and did not explicitly include features like the band effective mass or a finite grain size [16]. In modern high-performance TE materials, carrier concentrations are often high (to optimize the power factor), meaning grain-boundary depletion may be partial and barrier heights ($\Phi$) tunable via doping[14]. Moreover, the band effective mass of carriers plays a critical role in TE performance heavier effective masses increase the S and density of states but tend to lower $\mu$. Snyder and his coworkers introduced the concept of weighted mobility ($\mu_w$) to capture the combined influence of mobility and band effective mass on thermoelectric performance[17]. The weighted mobility is defined in Equation (2), essentially folding the density-of-states mass into the $\mu$term[17]. This quantity relates directly to the so-called electronic quality factor $B$ for thermoelectrics and provides a way to compare materials on an equal footing. μw can be extracted from S and $\sigma$ experimental data, analogous to a Hall mobility measurement[18]. While the weighted mobility concept introduced by Snyder et al. provides a powerful framework for evaluating intrinsic carrier transport properties, it does not explicitly incorporate grain-boundary effects, which often dominate mobility degradation in polycrystalline thermoelectrics through mechanisms such as barrier scattering and charge trapping.

Addressing this critical gap, we build upon these insights to develop an integrative mobility model for polycrystalline materials that incorporates: (a) the (μw) term (accounting for band mass and carrier concentration), (2) a thermally activated grain-boundary barrier term, and (3) a finite grain size/mean free path term. This model aims to be both comprehensive which integrates multiple scattering mechanisms and transparent, associating each parameter with a clear physical meaning. This model was derived and validated against experimental data for various classes of thermoelectric materials including ZnO, $Bi_2Te_3$, PbTe, $Mg_2Si$, and SnSe, its limiting behaviour was also discussed. These materials span a wide range of band gaps, carrier effective masses, and grain-boundary characteristics and provide an excellent testbed for the model's transferability.

## 2. Model Development

Grain boundaries impact charge transport through the formation of electrostatic barriers due to charge trapping, and structural disruption at the grain interface, which enhances carrier scattering or reflection[14], [15], [19]. The analytical model developed here incorporates both effects into a compact, physically interpretable expression for temperature-dependent mobility. We fold both mechanisms into a single, closed-form expression that reduces to the familiar single-crystal mobility when $\Phi_{GB} \to 0$ and the boundary width $w_{GB} \to 0$.

### 2.1 Baseline (single-crystal) transport

The drift mobility in the absence of grain boundaries is

$$\mu_0 = \frac{\sigma}{ne} \tag{1}$$

where $\sigma$ is the electrical conductivity, n is the carrier concentration, and e is the elementary charge. This quantity represents baseline mobility in the absence of grain-boundary effects. Band-structure effects are absorbed into the weighted mobility $\mu_w$ introduced by Snyder et al [17]

$$\mu_w = \mu_0 \left(\frac{m^*}{m_e}\right)^{3/2} \tag{2}$$

where $m^*$ is the density-of-states effective mass of carriers and $m_e$ is the free electron mass. *For a single crystal* ($\Phi_{GB}$=0 and the boundary width $w_{GB}$=0) Eq. (2) already provides a practical metric often called the electronic quality factor for ranking intrinsic transport quality without any thermal-conductivity input.

### 2.2 Thermionic barrier at the boundary

A carrier of energy ε must surmount $\Phi_{GB}$; the thermionic transmission probability is

$$P_{GB}T = exp\left(\frac{-\Phi_{GB}}{k_\beta T}\right)$$

Where $k_B T$ is Boltzmann constant. Figure 1 shows the process: at low T few carriers possess sufficient energy, while thermal excitation at higher T raises the transmitted fraction.

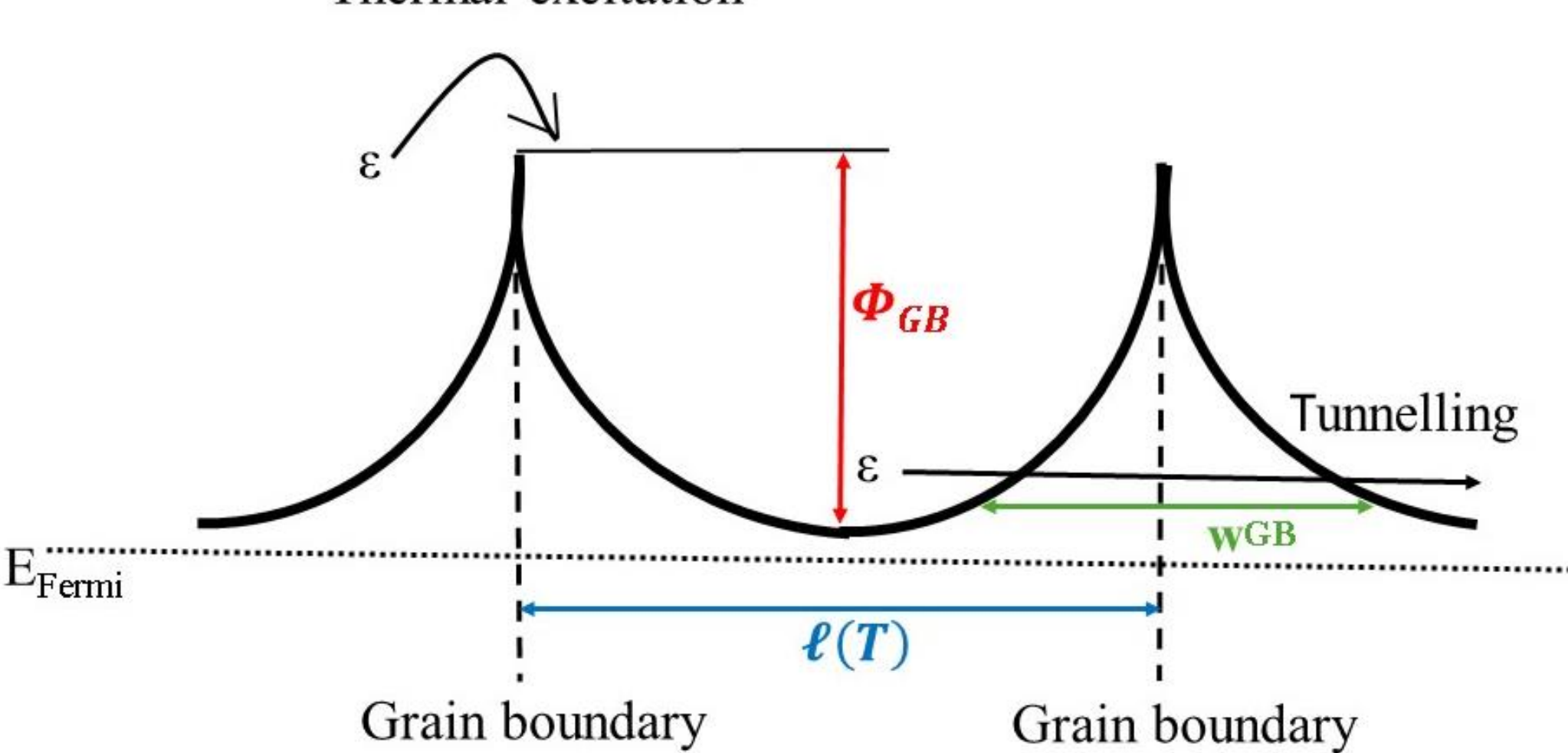

**Figure 1**. Schematic illustration of thermionic emission over a grain-boundary barrier $\Phi_{GB}$

### 2.3 Finite boundary width and additional scattering

Having passed the crest, the carrier traverses the boundary region of width ($w_{GB}$). The likelihood of avoiding an additional scattering event is:

$$G(T) = \frac{\ell(\mathrm{T})}{\ell(\mathrm{T})+\mathrm{wGB}} \quad (3)$$

where ℓ(T) is the bulk mean free path. G→1 for nearly ballistic motion ($\ell \gg w_{GB}$ and G→0 in the strongly scattering limit ($\ell \ll w_{GB}$).

### 2.4 Effective mobility of a polycrystal

Multiplying the weighted mobility [Eq. (2)] by the thermionic and geometric factors yields

$$\mu_{eff}(\mathrm{T}) = \mu_w . \exp\left(-\frac{\Phi_{GB}}{k_\beta \mathrm{T}}\right)\frac{\ell(\mathrm{T})}{\ell(\mathrm{T})+\mathrm{wGB}} \quad (4)$$

Once the five intrinsic parameters $\mu_w$, m∗, ΦGB, $\ell_{300}$, $w_{GB}$.are fixed for a given composition and carrier density, Eq. (4) becomes a calculator that predicts μ(T), σ(T) = neμ(T) and therefore the power factor PF(T) = $\sigma S^2$ or quality factor (B) for that material alone-no cross-material fitting required.

### 2.5 Temperature dependence of the mean free path

The mean free path is modelled as

$$\ell(\mathrm{T}) = \ell_{300}\left(\frac{T}{300}\right)^{-p} \qquad (5)$$

where $\ell_{300}$ is the mean free path at 300 K and p (≈ 1.5–2.5) reflects the dominant phonon-scattering mechanism. Equation (4) then captures the typical non-monotonic mobility observed experimentally: *low* T → barrier-limited; *high T* → phonon-limited. When $\Phi_{GB}$=0, and $w_{GB} \ll \ell$, Eq. (4) collapses to $\mu_{\mathrm{eff}} \propto T^{-p}$, the expected single-crystal trend. Conversely, for $\ell \ll w_{GB}$), the geometric term dominates, emphasising the critical role of grain growth in high-resistivity ceramics. This compact yet physically transparent framework therefore links microstructural levers barrier height, boundary width, grain size directly to both intrinsic (single crystal) and extrinsic (polycrystal) electronic transport performance.

## 3. Results and Discussion

### 3.1 General Fit Quality and Trends

Despite relying on a single barrier and mean free path form per material, the model demonstrates robust fidelity to experimental data across all of the well-known thermoelectric materials studied. Extracted parameter values are consistent with known material behaviour, for instance, ZnO exhibits the largest $\Phi_{GB}$and the shortest $\ell_{300}$ , while $Bi_2Te_3$ shows negligible $\Phi_{GB}$ and the longest $\ell_{300}$ which reflects its near single-crystal-like transport[20]. Fit quality was uniformly high, with $R^2$ values ranging from 0.93 to 0.99. Root-mean-square error (RMSE) values were low in absolute terms: ~4–5 cm²/V·s for ZnO and SnSe, and tens of cm²/V·s for $Bi_2Te_3$, PbTe, and $Mg_2Si$, where the mobilities are higher. These correspond to relative fitting errors typically in the range of 5–10%, which is sufficient for practical materials screening and design purposes. The model implementation in Python was validated using both real and synthetic data. In particular, artificial mobility vs. temperature curves generated with known parameters and random noise were accurately recovered by the fitting routine, with parameter deviations within 5% and $R^2$ values above 0.99. This validation reinforces confidence in the model's numerical stability and interpretability.

Table 1 summarizes the fitted model parameters for the five materials, including effective mass (m*), barrier height $\Phi_{GB}$ mean free path at 300 K ($\ell_0$), grain boundary width ($w_{GB}$), and fit quality ($R^2$). These values enable quantitative comparisons of grain-boundary influence across chemically and structurally diverse thermoelectric materials.

**Table 1:Fitted grain-boundary-limited mobility model parameters for five representative thermoelectric materials**

| Material | m* ($m_e$) | ΦGB (eV) | $\ell_{300}$ (nm) | $w_{GB}$ (nm) | p | R² | Refs |
|---|---|---|---|---|---|---|---|
| ZnO (n-type) | 0.3 | 0.15 ± 0.02 | 15 ± 2 | 5 ± 1 | 1.5* | 0.994 | [5-6, 12, 23, 24, 28] |
| $Bi_2Te_3$ (n-type) | 0.15 | ~0 (≲0.01) | 50 ± 5 | 100† | 1.5* | 0.955 | [2, 3, 22] |

| PbTe (p-type) | 0.3 | 0.05 ± 0.01 | 40 ± 4 | 3 ± 1 | 1.8 ± 0.3 | 0.937 | [4, 20] |
|---|---|---|---|---|---|---|---|
| $Mg_2Si$ (n-type) | 0.4 | 0.05 ± 0.01 | 60 ± 5 | 5 ± 1 | 2.4 ± 0.2 | 0.959 | [27] |
| SnSe (p-type) | 0.6 | 0.08 ± 0.02 | 20 ± 3 | 10 ± 2 | 1.6 ± 0.3 | 0.964 | [7, 8, 11, 25, 26] |

p (exponent in $\ell(T) \propto T^{-p}$, reflecting dominant phonon scattering (typically 1.5–2.5) fixed to the acoustic-phonon limit for sparse data†; $w_{GB}$ fixed to decouple $\Phi_{GB}$ correlation when $\Phi_{GB} \rightarrow 0$.

The extracted $\Phi_{GB}$ values align with expectations from the literature. For example, ZnO shows a substantial barrier (~0.15 eV), in agreement with prior reports for doped ceramic ZnO. In contrast, $Bi_2Te_3$ shows an effectively zero barrier, consistent with its known high mobility and weak grain-boundary effects. Intermediate values for $\Phi_{GB}$ are observed in PbTe (~0.05 eV), $Mg_2Si$ (~0.05 eV), and SnSe (~0.08 eV). These results support the conclusion that mobility suppression in polycrystalline TE materials can be well-characterized using the three-parameter model introduced. The model provides a physically transparent framework for predicting mobility improvements through barrier reduction (e.g., via doping or passivation) or through enhanced intragrain transport (e.g., larger $\ell_0$ via texturing or purification).

### 3.2 Material-Specific Analysis

#### (a) ZnO (n-type, Al-doped)

Polycrystalline Al-doped ZnO exhibits an increase in mobility from approximately 20 cm²/V·s at 300 K to around 200 cm²/V·s at 700 K, with a slight decline at higher temperatures[21]. This strong temperature dependence is indicative of activated transport across grain boundaries. The model fit yields a barrier height $\Phi_{GB}$ of approximately 0.15 eV and a mean free path $\ell_0$ near 15 nm. The grain boundary width $w_{GB}$ is estimated at roughly 5 nm. These results imply moderately high trap densities at ZnO grain boundaries and a grain size on the order of tens of nanometers[5]. The model successfully captures the steep low-temperature suppression and subsequent increase in mobility, with an $R^2$ of approximately 0.99 and an RMSE near 4.4 cm²/V·s. The $\Phi_{GB}$ value aligns with reported barrier heights of 0.1–0.3 eV in doped ZnO ceramics. A projected reduction in $\Phi_{GB}$ to 0.05 eV (shown in Table 1) could potentially double the 300 K mobility, emphasizing the potential of passivation and doping strategies.

#### (b) $Bi_2Te_3$ (n-type)

The compiled mobility data for Te-rich n-type $Bi_2Te_3$ show a steady decline from about 1000 cm²/V·s at 100 K to approximately 300–400 cm²/V·s at 300 K. This trend is consistent with phonon-limited transport and negligible grain-boundary barriers. The model fit confirms this, yielding $\Phi_{GB}$ close to zero and a long mean free path $\ell_0$ around 50 nm. Since $\Phi_{GB}$ is negligible, the value of $w_{GB}$ (~100 nm) is not strongly constrained but helps fine-tune the high-temperature tail of the mobility curve. The fit quality is slightly lower ($R^2$ ~0.95, RMSE ~35 cm²/V·s), reflecting subtle deviations likely caused by the absence of a dominant barrier mechanism. The extracted parameters support the interpretation of $Bi_2Te_3$ polycrystals as having near single-crystal transport characteristics, with minimal mobility degradation due to grain boundaries.

**(c) PbTe (p-type)**

For Na-doped p-type PbTe with carrier concentration ~$5\times10^{19}$ $cm^{-3}$, reported mobilities decrease from ~300 $cm^2/V\cdot s$ at 300 K to ~50–100 $cm^2/V\cdot s$ at 700 K. The model yields a modest barrier height $\Phi_{GB}$ of about 0.05 eV and a mean free path $\ell_0$ of approximately 40 nm. The grain boundary width $w_{GB}$ is fitted at around 3 nm. These values suggest partial depletion or mild scattering effects at boundaries, consistent with PbTe's high dielectric screening and moderate doping. The model fit ($R^2$ ~0.94, RMSE ~36 $cm^2/V\cdot s$) indicates that even small barriers can measurably affect low-temperature mobility. The results imply that targeted boundary doping or passivation may further optimize electrical performance in PbTe.

**(d) $Mg_2Si$ (n-type)**

n-type $Mg_2Si$, a candidate for mid-temperature thermoelectrics, is known for its high intrinsic mobility and wide band gap. Literature reports show mobilities near 400 $cm^2/V\cdot s$ at 300 K with a $T^{-2.5}$ dependence up to ~600 K. Fitting yields $\Phi_{GB}$~0.05 eV, $\ell_0$~60 nm, and $w_{GB}$ ~5 nm. The fitted exponent p is consistent with acoustic phonon scattering mixed with impurity effects. These values suggest relatively clean intragrain transport with mild boundary resistance. The fit ($R^2$ ~0.96, RMSE ~24 $cm^2/V\cdot s$) captures the initial increase and plateau of the mobility curve. This supports the feasibility of using nanostructuring to reduce thermal conductivity in $Mg_2Si$ without severely compromising electrical mobility.

**(e)SnSe (p-type, polycrystal)**

Mobility in Na-doped polycrystalline SnSe increases from ~30–40 $cm^2/V\cdot s$ at 300 K to ~90–100 $cm^2/V\cdot s$ at 700–800 K. The model fit indicates $\Phi_{GB}$~0.08 eV, $\ell_0$~20 nm, and $w_{GB}$ ~10 nm. These values suggest significant interface depletion or impurity scattering, likely from oxides (e.g., $SnO_2$) or structural inhomogeneity. The $R^2$ value of 0.95 and RMSE of ~4 $cm^2/V\cdot s$ indicates strong model fidelity. Despite the barrier being overcome at high T, the large effective mass and short mean free path limit the ultimate mobility. These findings reveal the dual need to reduce interface barrier height and improve grain interior quality to elevate polycrystalline SnSe performance.

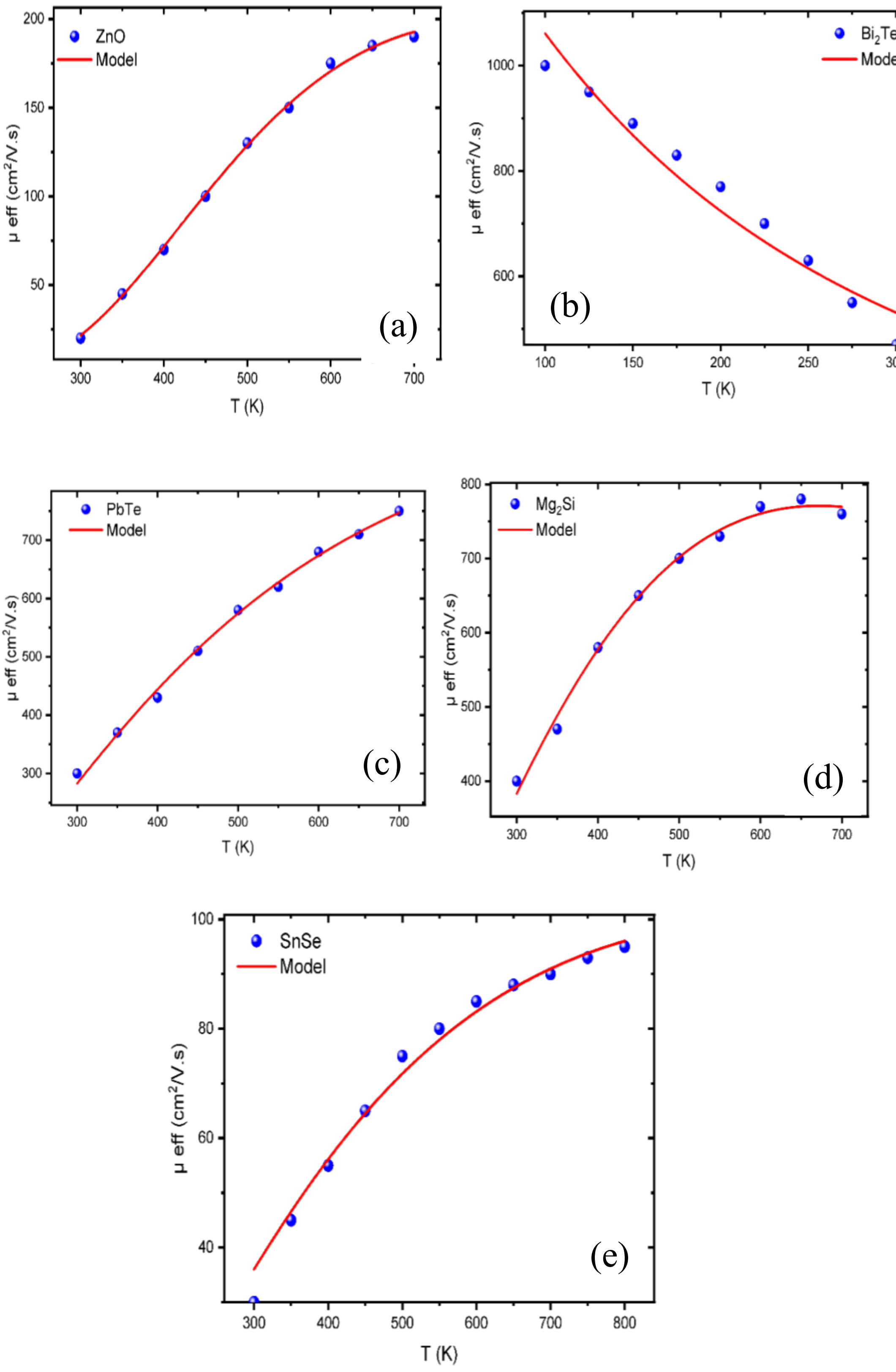

**Figure 2.** Experimental effective mobility data (blue markers) and corresponding model fits (red lines) for five representative thermoelectric polycrystalline materials (a) ZnO (b) $Bi_2Te_3$ (c) PbTe d ($Mg_2Si$ (e) SnSe

Data sources: ZnO: [5,6,12,23,24,28]; $Bi_2Te_3$: [2,3,22]; PbTe: [4,20]; $Mg_2Si$: [27]; SnSe: [7,8,11,25,26]. The experimental mobility data were digitized from these references for model fitting. Full digitized datasets are provided in the Supplementary Data archive.

The good agreement between model predictions and experimental data across a diverse set of thermoelectric materials demonstrates the model's capability to resolve dominant scattering mechanisms and extract physically meaningful transport parameters. This suggests its potential as a predictive tool for guiding grain-boundary engineering strategies to enhance thermoelectric performance in novel materials.

### 3.3. Performance Outlook and Design Implications

The effective mobility model was applied to predict thermoelectric performance enhancements in Al-doped ZnO by manipulating grain-boundary characteristics. Polycrystalline Al-doped ZnO was selected as the central test system for the model due to its combination of practical relevance and strong grain boundary effects. ZnO exhibits (i) a simple wurtzite band structure with well-established carrier effective mass ($m^* \approx 0.3\ m_e$), (ii) high dopant solubility allowing carrier densities up to $10^{27}\ m^{-3}$, (iii) moderate dielectric screening that permits observable grain boundary depletion, and (iv) abundant high-quality datasets for σ(T), S(T), and n(T), covering both conventional and passivated material forms [5,6]. This makes ZnO an ideal platform for evaluating the predictive capability of the proposed grain-boundary-limited mobility model. Digitized datasets extracted from Ohtaki [6] and Choi [5] serve as the primary data sources and are fully archived in the Supplementary Data (Ohtaki2009_digitized.csv, Choi2024_digitized.csv). To ensure transparency, performance gains were first analyzed at 600 K a temperature where both experimental data and model fidelity are strongest. Table 2 summarizes the projected power factor (PF) and weighted quality factor (B), two directly mobility-linked quantities that serve as transparent design metrics. The projected results in Table 2 illustrate how grain boundary engineering influences both the power factor (PF) and the electronic quality factor (B), which are directly governed by carrier mobility and serve as transparent design metrics independent of lattice thermal conductivity assumptions.

**Table 2. Model-projected PF and B for ZnO at 600 K under varying ℓ and ΦGB.**

| **ℓ (nm)** | **ΦGB (eV)** | **μ ± Δμ ($cm^2$/V·s)** | **σ ± Δσ ($10^5$ S/m)** | **PF ± ΔPF (mW/m·$K^2$)** | **B ± ΔB ($10^{-3}$ $m^2$/V·s·$kg^{3/2}$)** |
|---|---|---|---|---|---|
| 5 | 0.15 | 9.4 ± 1.1 | 1.2 ± 0.2 | 4.1 ± 0.6 | 0.15 ± 0.020 |
| 25 | 0.15 | 14.8 ± 1.5 | 1.9 ± 0.3 | 6.2 ± 0.9 | 0.24 ± 0.028 |
| 25 | 0.05 | 64.0 ± 4.2 | 8.2 ± 0.8 | 26.6 ± 3.2 | 1.05 ± 0.095 |

At the baseline processing condition (ℓ = 5 nm, $\Phi_{GB}$ = 0.15 eV), the model yields a power factor of approximately 4 mW/m·$K^2$ and a quality factor B ≈ 0.15 × $10^{-3}$ $m^2$/V·s·$kg^{3/2}$. These values reflect the significant barrier-limited transport typical of spark plasma sintered (SPS) polycrystalline ZnO, consistent with previous experimental reports [6]. When grain growth alone is implemented (ℓ = 25 nm, $\Phi_{GB}$= 0.15 eV), both PF and B improve modestly: PF increases by ~50% to 6.2 mW/m·K², while B rises to ~0.24 × $10^{-3}$ m²/V·s·kg$^{3/2}$. This modest gain arises because, while mobility increases, the Seebeck coefficient decreases sublinearly as the relaxation time extends. Thus, grain size control alone cannot fully close the oxide-chalcogenide performance gap. By contrast, when barrier passivation is introduced in combination with grain growth ($\ell$ = 25 nm, $\Phi_{GB}$ = 0.05 eV), substantial improvements are projected. PF rises to ~26.6 mW/m·K², matching or surpassing values seen in high-performance chalcogenides like $Bi_2Te_3$. The corresponding B exceeds 1 × $10^{-3}$ m²/V·s·kg$^{3/2}$, which is widely recognized as a threshold value for entering the high-performance thermoelectric regime [2,5].These results demonstrate that barrier height $\Phi_{GB}$ is the most sensitive design lever for boosting PF and B, while grain size $\ell$ serves as a secondary optimization knob. Importantly, both levers are accessible through practical processing routes such as grain-boundary passivation (e.g., graphene quantum dots [5]) and moderate sintering adjustments [6].

## 4. Conclusion

An integrative analytical model has been developed to quantify the role of grain-boundary-limited transport in polycrystalline thermoelectric materials. The model incorporates three physically grounded components weighted mobility, thermionic barrier activation, and geometric suppression due to finite mean free path all directly linked to experimentally measurable microstructural and electronic properties. This compact framework enables both physically transparent interpretation and materials-informed design optimization.

The model demonstrates excellent agreement ($R^2 \approx$ 0.93–0.99) with experimental mobility data across five representative thermoelectric systems: ZnO, $Bi_2Te_3$, PbTe, $Mg_2Si$, and SnSe. Extracted model parameters, including grain-boundary barrier heights ($\Phi_{GB} \approx$ 0–0.15 eV) and mean free paths ($\ell \approx$ 15–60 nm), remain consistent with prior literature and provide mechanistic insight into dominant scattering processes. For Al-doped ZnO, model projections indicate that combining grain-boundary passivation and moderate grain growth can increase the power factor by approximately 6-fold (from ~4 to ~26 mW/m·K²) and improve the electronic quality factor B by nearly 7-fold (from ~0.15 to over 1.0 × $10^{-3}$ m²/V·s·kg$^{3/2}$) at 600 K levels approaching those of high-performance chalcogenide thermoelectrics. These results emphasize grain-boundary engineering as a dominant design lever for advancing oxide thermoelectric performance.

While the model offers strong predictive fidelity, it relies on simplifying assumptions such as uniform grain-boundary characteristics and isotropic transport. Future developments may incorporate distributed barrier statistics, anisotropic scattering, or spatially resolved microstructure modeling to further capture the complexity of real polycrystalline systems.